# Enhancing Data Completeness in Time Series: Imputation Strategies for Missing Data Using Significant Periodically Correlated Components

Asmaa Ahmad[1]*, Eric J Rose[1], Michael Roy[2], Edward Valachovic[1]

[1] Department of Epidemiology and Biostatistics, College of Integrated Health Sciences, University at Albany, State University of New York, One University Place, Rensselaer, NY

[2] NYS Department of Health

*Corresponding author.

E-mail: Aahmad4@albany.edu (AA)

**Abstract:**

Missing data is a pervasive issue in statistical analyses, affecting the reliability and validity of research across diverse scientific disciplines. Failure to adequately address missing data can lead to biased estimates and consequently flawed conclusions. In this study, we present a novel imputation method that leverages significant annual components identified through the Variable Bandpass Periodic Block Bootstrap (VBPBB) technique to improve the accuracy and integrity of imputed datasets. Our approach enhances the completeness of datasets by systematically incorporating periodic components into the imputation process, thereby preserving key statistical properties, including mean and variance. We conduct a comparative analysis of various imputation techniques, demonstrating that our VBPBB-enhanced approach consistently outperforms traditional methods in maintaining the statistical structure of the original dataset. The results of our study underscore the robustness and reliability of VBPBB-enhanced imputation, highlighting its potential for broader application in real-world datasets, particularly in fields such as healthcare, where data quality is critical. These findings provide a robust framework for improving the accuracy of imputed datasets, offering substantial implications for advancing research methodologies across scientific and analytical contexts. Our method not only impute missing data but also ensures that the imputed values align with underlying temporal patterns, thereby facilitating more accurate and reliable conclusions.

# 1. Background:

Missing data present a significant challenge to statistical analyses, complicating inference and reducing accuracy across many applications (Rubin, 1976). Missingness can introduce biases that obscure genuine patterns and trends, particularly periodic structures, thereby compromising research validity and predictive performance (Schafer & Graham, 2002). The approach discussed here specifically enhances the quality of imputed values for datasets with periodic characteristics, making it especially valuable for time series analysis. However, it is important to note that this method is most effective for data characterized by periodic behavior and may not provide the same benefits for datasets without such patterns. Addressing these challenges enhances the reliability of statistical modeling and improves insights across diverse domains.

This study addresses the issue of missing data by utilizing the Variable Bandpass Periodic Block Bootstrap (VBPBB) method (Valachovic, 2024). Unlike traditional imputation techniques, VBPBB specifically extracts and leverages significant periodic components, aligning the imputation process with the inherent structure of the data. This method extends the approach of Cleveland et al. (1990) by emphasizing the preservation of temporal patterns, which is critical for accurately modeling the cyclic behavior often observed in health-related time series data.

# 2. Overview of the Imputation Process

To provide a structured framework for understanding the methodology, Figure 1 presents a high-level flowchart outlining the major stages of the data imputation process. The flowchart illustrates the sequential progression from the simulation of missing data, to the extraction of significant periodic components using the Variable Bandpass Periodic Block Bootstrap (VBPBB) method, followed by the integration of these components into the Amelia II package for multiple imputation. After imputation, smoothing techniques are applied to refine the reconstructed datasets. The process concludes with an evaluation of imputation performance using Mean Absolute Error (MAE) and Root Mean Squared Error (RMSE) metrics. A detailed description of each step is provided in the subsequent sections.

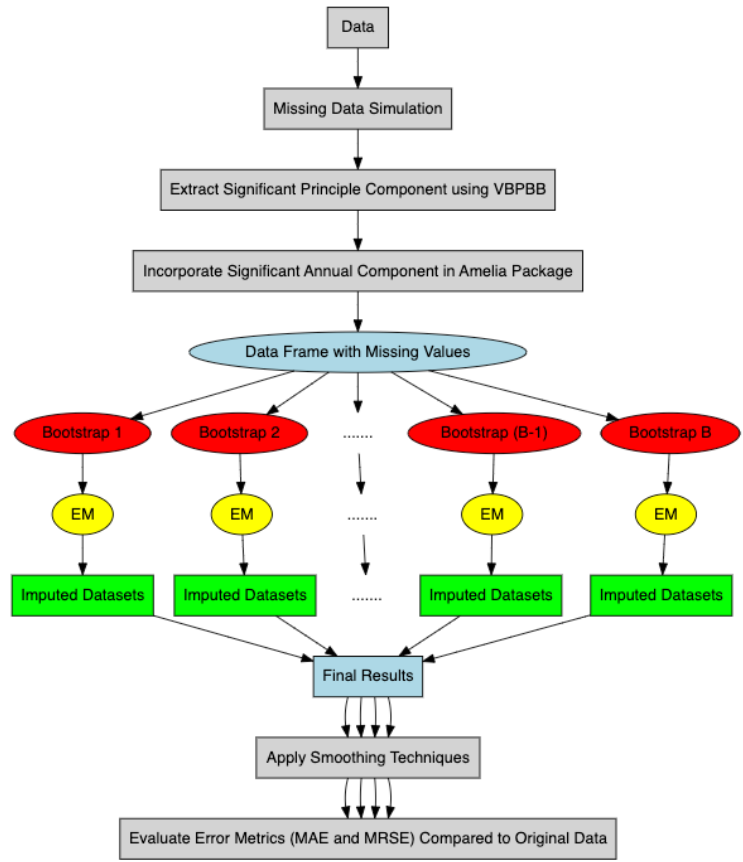

**Figure 1:** Flowchart of Data Imputation Process Using VBPBB

## 3. Impact of Missing Data in Research

Missing data is a pervasive challenge in research that significantly impacts the integrity and validity of study results across various disciplines. The issue arises in nearly all studies and profoundly influences the conclusions that can be reliably drawn. As Bennett (2001) highlights, the presence of missing data beyond a 10% threshold typically introduces biases that can skew analysis, disrupting the underlying data structure, reducing statistical power, and leading to biased parameter estimates and incorrect standard errors. These complications render research findings unreliable and potentially misleading.

To address this challenge, various methods have been developed for handling missing data, one of which is multiple imputation (MI), a widely recognized and robust approach currently supported by many statistical software packages. In this method, multiple values are imputed to replace each missing value, accounting for the uncertainty associated with the true value. Auxiliary variables—variables correlated with the variables of interest yet not part of the specified analysis—can be incorporated into the MI process to increase the validity of the imputed data (Sinharay et al., 2001; Madley-Dowd et al., 2019). The method begins by predicting missing data using information from other observed variables (Sinharay et al., 2001). These predicted values are then used to fill in the missing observations, creating a complete dataset known as the imputed dataset. Multiple imputed datasets are generated, analyzed separately, and then combined to produce a single overall analysis result. This process ensures that the inherent uncertainty in missing data is accounted for, leading to more reliable statistical inferences, and reducing the risk of biased estimates.

When the proportion of missing data is very small, however, the benefits of multiple imputation may be limited. Schafer (1999) suggests that when the missing data rate is below approximately 5%, the improvements offered by multiple imputation over simpler methods are often negligible. Historically, in cases where missing data were substantial, researchers recommended restricting analyses solely to subjects with complete data—an approach known as "complete case" analysis (Austin et al., 2021). This method operates under the assumption that attempting to impute large gaps might distort the original data structure and introduce additional inaccuracies.

Despite these traditional views, more recent research, including a study by Madley-Dowd et al. (2019), challenges the notion that imputation should be avoided when missingness is extensive. Their results demonstrate that advanced imputation techniques can effectively reduce biases and preserve analytical integrity even when up to 70% of the data is missing. This shift highlights the growing emphasis on employing sophisticated methods for handling missing data to ensure the accuracy and reliability of research outcomes.

## 4. Types of Missing Data

Understanding the mechanisms behind missing data is essential for selecting the appropriate methods to address it, as categorized by Rubin (1976):

**4.1 Missing Completely at Random (MCAR)**

Missing Completely at Random (MCAR) occurs when the probability of missing data is entirely independent of both observed and unobserved variables, meaning that missingness happens purely by chance without any systematic pattern (Jamshidian & Jalal, 2010). Under this assumption, missing data do not introduce bias, making MCAR the least problematic type of missingness in statistical analysis. However, testing for MCAR is essential before selecting an appropriate method for handling missing data. One widely used approach is the methodology proposed by Jamshidian and Jalal (2010), which compares covariance structures across groups with identical missing data patterns. Their method involves imputing missing values using either a normality-based approach, which assumes a multivariate normal distribution, or a distribution-free approach, which does not rely on specific distributional assumptions. The equivalence of covariance matrices between these groups is then tested using Hawkins' test for normally distributed data or non-parametric methods when normality is not assumed. Significant differences in covariance structures suggest that the data are not MCAR, meaning the missingness is related to observed or unobserved variables, classifying the data as Missing at Random (MAR) or Missing Not at Random (MNAR).

Common approaches for handling MCAR data include listwise deletion, pairwise deletion, and mean imputation. Listwise deletion (complete case analysis) is a straightforward method where only cases with no missing values on the variables of interest are retained. While this approach ensures unbiased estimates under the MCAR assumption, it significantly reduces the sample size and may introduce bias if the data are not truly MCAR (Donner, 1982). Pairwise deletion (available case analysis) preserves more data by including all available observations for each specific analysis rather than removing entire cases. When data are MCAR, pairwise deletion produces consistent estimates of parameters that can be expressed as functions of population means, variances, and covariances (Glasser, 1964). However, like listwise deletion, if the data are not MCAR, pairwise deletion may yield biased estimates (Allison, 2009). Mean imputation replaces missing values with the overall estimated mean of the variable, maintaining a complete dataset for analysis. However, this approach does not introduce new information, can distort variability, and may underestimate standard errors, leading to biased parameter estimates when missingness is not purely random (Bennett, 2001). While these methods can effectively address MCAR data without introducing systematic bias, they often reduce the effective sample size, potentially impacting statistical power (Bennett, 2001).

**4.2 Missing at Random (MAR)**

MAR occurs when the probability of missing data is systematically related to other observed variables but remains independent of the missing values themselves. This type of missingness implies a structured pattern that can be predicted based on observed data. As a result, methods such as Multiple Imputation (MI) and Maximum Likelihood (ML) estimation are effective for MAR, as they leverage the relationships within the observed data to impute missing values without introducing significant bias (Enders, 2010; Graham, 2009). MI, in particular, involves creating multiple plausible datasets and averaging the results to enhance the robustness of the estimates, while ML utilizes all available data to estimate parameters that best fit the observed data structure. The structured nature of MAR allows for reliable prediction of missing values, preserving the relationships between variables within the dataset.

**4.3 Missing Not at Random (MNAR)**

MNAR is the most complex type of missingness, where the probability of missingness is dependent on unobserved values, making the missing data mechanism inherently related to the missing values. This situation presents a significant challenge, as the missing data is informative and directly related to the unobserved values themselves (Rubin, 1976). Standard imputation techniques are generally inadequate for MNAR, as they fail to account for the bias that the MNAR mechanism introduces. Advanced approaches, such as selection models and pattern-mixture models, are necessary to manage MNAR data. These models explicitly account for the missingness mechanism during analysis, though they require robust assumptions and complex modeling frameworks to address the bias inherent in MNAR data (Little & Rubin, 2019).

## 5. Practical Assessment and Management of Missing Data

A thorough analysis of the pattern of missingness within a dataset is crucial before implementing any imputation techniques. Such analysis not only helps assess the complexity of the missing data problem but also influences the choice of imputation method, thereby minimizing potential biases.

While the proportion of missing data is an important consideration, it is not the sole factor researchers should assess. According to Tabachnick and Fidell (2012), the causes and patterns of missing data significantly impact research outcomes, often more so than the proportion of missing data alone. This view is supported by Madley-Dowd et al. (2019), who demonstrated that the mechanism of missingness should guide the choice of imputation method. Their findings suggest that multiple imputation (MI) can produce valid and reliable results even when up to 50% of data is missing, provided the missing data is classified as MAR. This challenges traditional views that high proportions of missing data are inherently problematic and underscores the importance of a detailed approach to data imputation. By understanding both the quantity and the nature of the missing data, researchers can effectively mitigate bias and maintain analytical integrity using appropriate techniques.

## 6. Methodology:

This study addresses the complexities of managing missing data in time series analysis, crucial for ensuring the accuracy and reliability of studies. Traditional imputation methods often fail to adequately preserve the inherent temporal dynamics of time series data, potentially leading to substantial analytical deficiencies. Khan and Lazar (2023) discussed that numerous efforts have been directed towards developing effective strategies to mitigate these issues. However, conventional approaches often require extensive knowledge about the underlying data and struggle with adaptability across different data types, lacking the capacity to generalize to new, unseen datasets. Furthermore, these methods typically overlook the relational attributes like spatial and temporal components, essential for enhancing prediction accuracy and reliability in subsequent analyses.

In response to these challenges, our study introduces an advanced imputation technique that combines the Variable Bandpass Periodic Block Bootstrap (VBPBB) with Amelia II package. This integration leverages the strengths of both approaches: VBPBB's effectiveness in handling periodic and seasonal variations and Amelia II's robustness in broader statistical applications. This innovative strategy promises to enhance the handling of missing data, providing a more robust framework for maintaining the integrity of time series datasets.

**6.1 Variable Bandpass Periodic Block Bootstrap (VBPBB):**

The Variable Bandpass Periodic Block Bootstrap (VBPBB) method, introduced by Valachovic (2024), is designed to analyze time series data characterized by significant periodic components, particularly in cases where traditional methods struggle with inherent instabilities. The VBPBB method and its extension the VMBPBB for multiple periodic components introduced by Valachovic (2025) integrates concepts from Zurbenko's (1986) iterated moving average bandpass filters to isolate specific frequency components within the data, ensuring a refined and noise-reduced analysis. A key feature of VBPBB is the incorporation of a bandpass filter that targets a specific principal component (PC) frequency, isolating variations at or near the chosen frequency while attenuating those outside the band. This selective filtering enhances the clarity and accuracy of periodic data analysis by reducing the impact of shocks, interventions, noise, outliers, and long-term trends, allowing a sharper focus on essential periodic patterns. By reconstructing a PC time series that retains only the critical correlation structure of a single PC component, VBPBB enables a more precise and structured analysis of periodic data.

Block bootstrapping is a fundamental aspect of VBPBB, ensuring that the correlation structure of the time series is preserved while resampling. By selecting a block size that matches the period of the PC component, VBPBB maintains temporal integrity and improves estimation precision. This results in smaller confidence interval (CI) sizes for the periodic mean compared to traditional periodic block bootstrap methods, as supported by Brockwell and Davis (2002), who emphasize the importance of maintaining dependency structures in time series resampling. Furthermore, spectral analysis, particularly through periodograms, plays a crucial role in identifying dominant periodic components, enhancing the method's effectiveness in extracting relevant patterns. The collaborative work of Valachovic and Shishova (2024) highlights how periodograms aid in uncovering key periodic structures within complex datasets.

In practical applications, VBPBB isolates major periodic components such as annual or seasonal variations, preserving original correlation structures while ensuring a clearer interpretation of cyclic patterns. This selective filtering and resampling process significantly improves upon traditional resampling techniques, providing a more precise, data-driven, and methodologically rigorous approach to periodic time series analysis. By focusing exclusively on relevant frequency components, reducing noise interference, and preserving temporal dependencies, the VBPBB method offers a robust and flexible solution for capturing intricate temporal dynamics, making it an invaluable tool for researchers working with periodic time series data.

**6.2 KZFT Filters Application**

As part of the VBPBB implementation, the Kolmogorov-Zurbenko Fourier Transform (KZFT) is applied to principal component (PC) time series data to achieve the selective filtering necessary for isolating key spectral components. The KZFT method, which integrates elements of moving averages and convolution within the time domain (Zurbenko, 1986), decomposes a time series into its constituent frequency components by applying a series of weighted moving averages. This iterative smoothing process extracts frequency content information at multiple scales, allowing for precise analysis of periodic structures.

In VBPBB, the frequency center ($v$) for each KZFT filter is set to the reciprocal of the period ($1/p$), where p denotes the period of the PC component of interest. Additional parameters, such as the width of the moving average filter window ($m$) and the number of filter iterations ($k$), are carefully selected to ensure the filter passes only the desired PC component frequency (Yang and Zurbenko, 2007). When analyzing two or more PC components, strategically placed cutoff frequencies between PC component frequencies are used to enhance separation (Valachovic,

2025). This approach refines the dissection and analysis of periodicity, contributing significantly to the methodological advancements achieved through the VBPBB framework.

**6.3 Amelia II**:

Amelia II, developed by Honaker, King, and Blackwell (2011), is an R package designed for multiple imputation. This statistical technique addresses missing data within complex datasets, including those with time series and cross-sectional elements. Amelia II utilizes a bootstrapped version of the Expectation-Maximization (EM) algorithm, integrated with advanced statistical modeling, to effectively manage missing data characterized by non-response and attrition. It operates under the assumption, common to most multiple imputation methods, that data are missing at random (MAR). This means the pattern of missingness depends only on the observed data, $D^{obs}$, and not on the unobserved data, $D^{mis}$.

The EM algorithm in Amelia II involves:

- **Expectation Step (E-step)**: Calculates the probabilities of missing data points based on the observed data and current estimates of model parameters.
- **Maximization Step (M-step)**: Updates the parameters to maximize the likelihood of the data, integrating the imputed values into the dataset seamlessly.

Zhang (2016) describes the Bootstrap-based EM algorithm as a sophisticated method for handling missing data, ideal for generating multiple imputations. The process begins by creating multiple samples from the original dataset; specifically, it draws $m$ samples, each with n observations, where $n$ is the size of the original dataset. For each sample, the EM method calculates point estimates of the mean and variance. Using these estimates, missing values are then imputed by drawing from predictive distributions based on the estimated parameters, ensuring that each imputation appropriately reflects the uncertainty in the missing data (Honaker et al., 2011). These imputed values are inserted into the original dataset, resulting in $m$ complete datasets ready for further analysis.

Amelia II enhances imputation accuracy by utilizing bootstrapping, generating multiple samples of the original dataset through resampling with replacement and applying the EM algorithm to each. This not only improves the accuracy of imputations but also provides crucial estimates of variability, essential for robust statistical analysis.

Particularly skilled at handling time series data, Amelia II ensures that imputed values maintain essential temporal correlations, seasonal patterns, and trends. It incorporates ARIMA models within the EM algorithm for a nuanced treatment of datasets with inter-temporal dependencies. Moreover, the integration of significant periodic components through methods like the Variable Bandwidth Periodic Block Bootstrap (VBPBB) allows for adjustments for temporal and nonlinear patterns, enhancing the quality of imputation.

The flexibility of Amelia II enables efficient handling of large and complex datasets, accommodating variables not affected by missingness to refine the imputation process further. Coupled with the provision of uncertainty measures around point estimates, Amelia II stands out as a valuable tool for researchers dealing with incomplete data in time series analysis. By leveraging VBPBB for precise periodic identification, Amelia II enhances the imputation process by treating significant periodic components as auxiliary variables. This adjustment for temporal and nonlinear patterns significantly reduces the risk of bias introduced by inappropriate handling of time-related dependencies.

### 6.4 Application of Smoothing Techniques

To refine our analysis, various smoothing techniques, including LOESS smoothing, and moving averages were applied uniformly across the entire dataset after initial preprocessing, which adjusted for seasonality and trends. Each technique was chosen for its ability to reduce noise while preserving essential data characteristics.

**Moving Average:** This technique smooths out short-term fluctuations and highlights longer-term trends or cycles in time series data. By calculating the average of data points within a specified moving window, it effectively filters out 'noise' and clarifies the underlying trend (Box et al., 2015). This window can be centered, depending on the arrangement of the data points relative to the time of prediction (Isnanto, 2022). Moving averages are particularly useful for datasets with seasonality, as adjusting the window size to match the length of the seasonal pattern can help mitigate seasonal variations.

**LOESS (Locally Estimated Scatterplot Smoothing)** is a non-parametric method that uses local polynomial regression to smooth fluctuations in data. Isnanto (2022) discussed that LOESS can fit more flexible shapes because it uses a localized fitting technique, where each data point is smoothed by considering only the nearby points defined by a smoothing parameter. This parameter controls how much the surrounding points influence the smoothing, allowing LOESS to adapt to changes in the data's statistical properties over time.

In time series analysis, two popular smoothing techniques are filtering, often involving averaging, and local regression, as discussed by R. Rizal Isnanto (2011). Both methods utilize a concept called a span or window size, which specifies the range of neighboring data points considered for smoothing at each data point across the dataset. As this window of points moves through the dataset, it calculates the smoothed value for each point. Isnanto points out that the size of the span significantly impacts the results: a larger span results in smoother data but with reduced detail (resolution), whereas a smaller span provides finer details at the cost of smoothness. He emphasizes that determining the ideal span size generally involves trial and error, as it varies based on the characteristics of the dataset and the specific requirements of the smoothing method employed.

### 6.5 Evaluation Metrics for Imputation Methods: MAE and RMSE
For assessing the accuracy and reliability of our imputation methods, we also employed Mean Absolute Error (MAE) and Root Mean Squared Error (RMSE). These metrics measure the deviation of imputed values from actual observations, offering a quantitative basis for comparing the performance of different imputation techniques:

- **Mean Absolute Error (MAE):** calculates the average of the absolute differences between the predicted (or imputed) and actual values across all observations. A lower MAE indicates that the predictions are closer to the actual outcomes, demonstrating better performance of the imputation method.

  It is defined as:

  $$MAE = \frac{1}{n} \sum_{i=1}^{n} |y_i - \hat{y}_i|$$

  where $y_i$ represents the actual values, $\hat{y}_i$ denotes the predicted or imputed values, and $n$ is the number of data points. A lower MAE indicates better performance, showing that predictions are closer to actual outcomes (Willmott & Matsuura, 2005).

- **Root Mean Squared Error (RMSE):** provides a measure of the average magnitude of the errors, penalizing larger errors more heavily due to the squaring of the error terms. It is particularly useful when large errors are undesirable in the analysis.

  The formula for RMSE is:

  $$RMSE = \sqrt{\frac{1}{n} \sum_{i=1}^{n} (y_i - \hat{y}_i)^2}$$

  A lower RMSE indicates a closer fit of the model to the data, suggesting better performance of the imputation method (Chai & Draxler, 2014).

These metrics are critical not only for evaluating the immediate accuracy of the imputed data but also for ensuring that the methods used adhere to the expected standards of reliability and precision in statistical analysis.

## 7. Data simulation for Method Evaluation

To demonstrate the use of the Amelia II package, incorporating the Variable Bandpass Periodic Block Bootstrap (VBPBB), and to assess the efficacy of data imputation with VBPBB versus imputation methods without VBPBB, we conducted a simulation study within the Missouri Historical Agricultural Weather Database. Specifically, daily average temperature records were extracted from the Commercial Agriculture Automated Weather Station Network, maintained by the University of Missouri Extension (University of Missouri Extension, n.d.). Using a web-based interface, we selected Albany (Gentry County) as the location for our analysis, with the study period specified from January 1, 2014, to December 16, 2024.

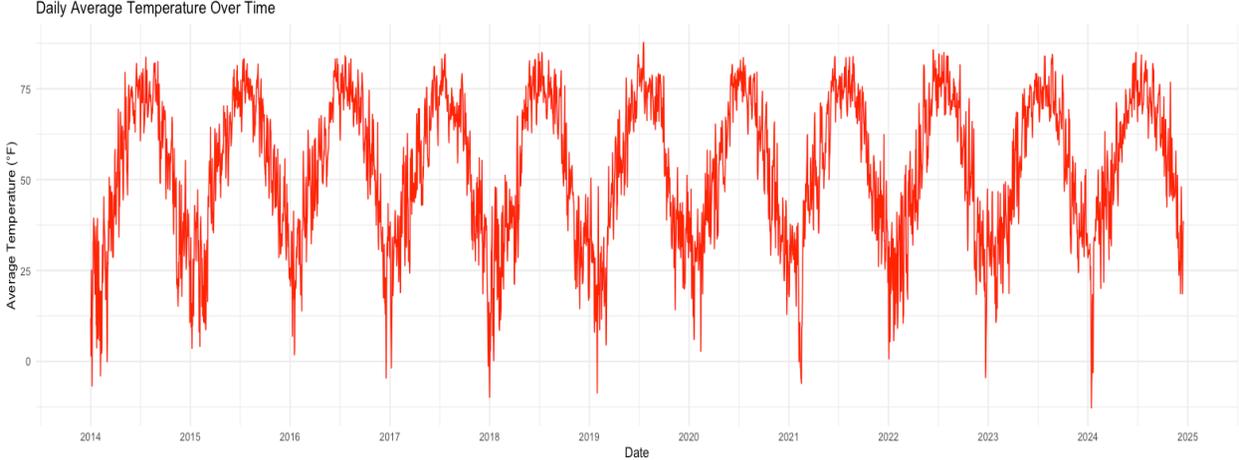

**Figure 2:** the daily average temperatures from January 2014 to December 2024, as recorded by the Commercial Agriculture Automated Weather Station Network in Albany (Gentry County), Missouri. The temperatures are expressed in degrees Fahrenheit and display clear seasonal fluctuations consistent with regional climatic patterns.

The simulation was conducted using R version 4.1.1. We categorized days into weekends and weekdays using the wday() function from the lubridate package, assigning 13% of the dataset as missing with a disproportionate number allocated to weekends (30%) to mimic higher non-response rates during these periods. Temperature values for these selected days were designated as missing (NA) through random selection using the sample() function.

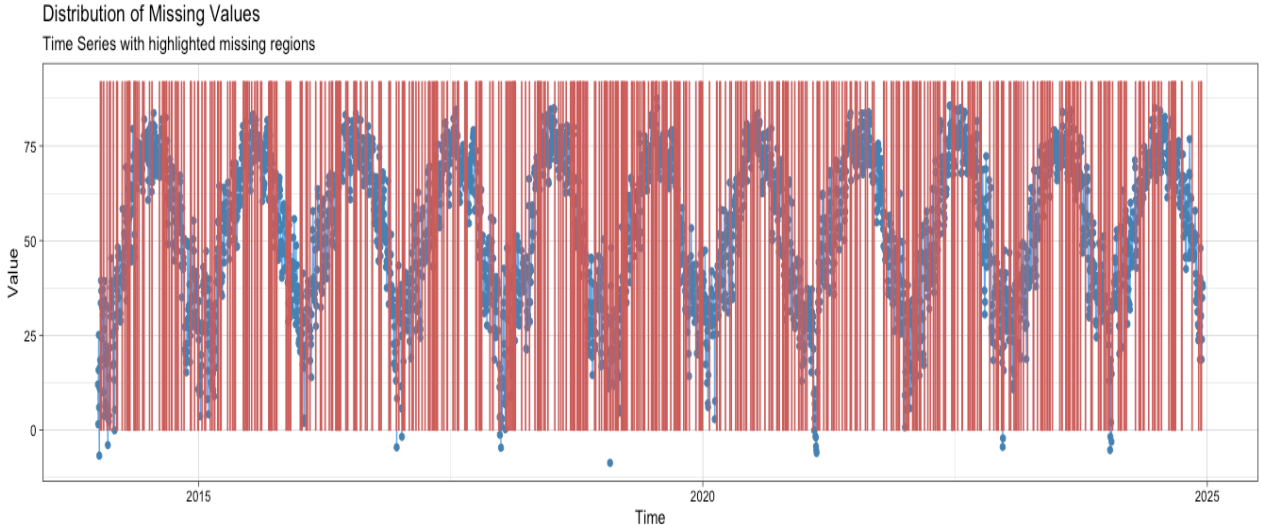

**Figure 3:** presents the distribution of missing values in a dataset of daily average temperatures from 2015 to 2025. The graph overlays red vertical bars on a time series plot to indicate days with missing temperature data, while observed temperature values are represented as blue dots.

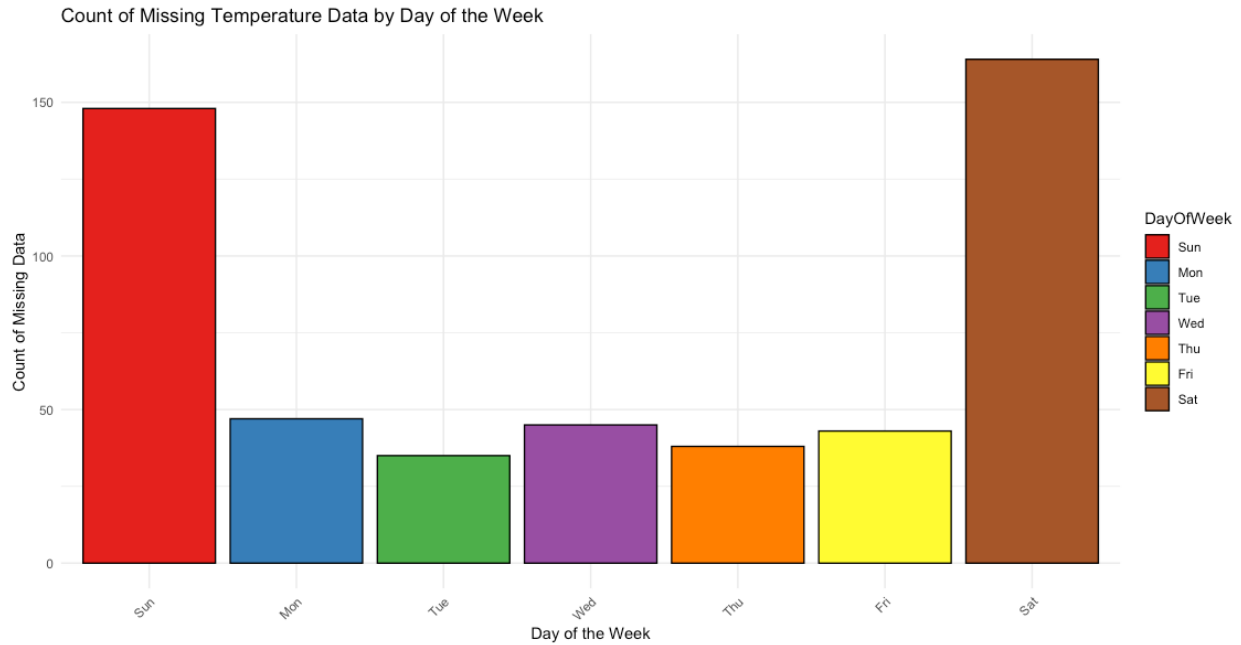

**Figure 4: Count of missing data across different days of the week**

Descriptive analysis of missing values is crucial as it helps understand the extent and pattern of missing data within a dataset, ensuring the reliability and validity of subsequent analyses. This involves identifying potential biases and determining the most appropriate strategies for handling missing values. Figure 4 illustrates the proportion of missing data across different days of the week, using distinct colors for each day from Sunday to Saturday. It reveals that a substantial amount of data is missing on weekends, particularly on Sundays and Saturdays, which together account for about 60% of the missing entries. This pattern created to introduce a systematic issue with data reporting or collection during weekends, likely influenced by reduced staffing or operational hours in facilities where the data is gathered. The consistent occurrence of missing data during these times indicates that the data is not missing completely at random (MCAR) but is instead linked to an external factor, such as operational schedules, which does not depend on the unobserved values themselves (i.e., it is missing not at random or MNAR). Given this context, we classify the missing data as Missing at Random (MAR). The missing data is dependent on the day of the week (an observed variable) and not on the unobserved data. This classification allows us to use imputation methods suitable for MAR data to handle the missing values effectively, ensuring robust and reliable statistical analysis. This insight necessitates the use of imputation methods that can account for these systematic missing data patterns to ensure the accuracy and reliability of subsequent analyses.

The extraction of significant periodic components from the temperature dataset using the Kolmogorov-Zurbenko Filter Transform (KZFT) function illustrates an approach for identifying principal components (PCs) associated with both natural and anthropogenic influences. The KZFT function, implemented via the kza package, applies bandpass filters to isolate dominant frequencies corresponding to seasonal and operational cycles, such as daily, weekly, and monthly patterns.

This methodology facilitates the differentiation between naturally occurring cycles, such as seasonal variations driven by atmospheric and climatic processes, and human-induced periodic patterns linked to work schedules, industrial activity, and energy consumption. By selectively filtering and extracting these components, KZFT provides a precise characterization of the underlying drivers of variability within the dataset.

Understanding both environmental and anthropogenic influences enhances the interpretability of time series data, particularly in climate and environmental studies, where distinguishing natural cycles from human-driven impacts is essential for accurate modeling and forecasting. By isolating periodic components at specific frequencies, KZFT enables a comprehensive analysis of temporal patterns, improving the ability to detect, quantify, and interpret seasonal and operational influences in complex datasets.

For the yearly component analysis, a window size of 731 days was used, targeting frequencies like 1/365, 2/365, 3/365, 4/365, 5/365, and 6/365, which capture not only the annual cycles but also biannual and other sub-annual trends. Monthly components were examined with a window size of 101 days at frequencies of 1/30, 2/30, and 3/30, while weekly patterns utilized a window size of 21 days, examining frequencies of 1/7, 2/7, and 3/7. This methodology allows the model to detect a broad spectrum of low-frequency components, extending beyond straightforward cyclic patterns to encompass higher-order harmonics, thereby accommodating more intricate seasonal variations.

To ensure the accuracy and relevance of this analysis, only statistically significant harmonics were retained based on a non-parametric bootstrap procedure. For each extracted periodic component, 95% confidence intervals (CIs) were constructed from the bootstrap samples generated using the Variable Bandpass Periodic Block Bootstrap (VBPBB) method. A periodic component was deemed statistically significant if the maximum lower bound of its confidence interval was greater than the minimum upper bound, indicating that the confidence interval for the periodic component amplitude did not contain zero. This selection criterion ensured that each frequency component contributed meaningful information to the dataset.

After extraction, the isolated principal components were resampled using VBPBB to preserve the integrity of their autocorrelation structures. Notably, the annual and weekly components were identified as statistically significant at the 95% confidence level in characterizing the periodic behavior of the temperature data. The prominence of the annual component aligns with expected seasonal variations driven by solar cycles, reinforcing its critical role in modeling long-term temperature trends. In contrast, the weekly component captures systematic short-term fluctuations likely associated with human or operational factors influencing data collection. Its statistical significance is further corroborated by the observed pattern of increased missing data on weekends, suggesting potential gaps due to reduced staffing, maintenance schedules, or reporting delays. Incorporating this weekly signal into the model enhances the imputation process by preserving short-term temporal dependencies, thereby improving the overall fidelity and robustness of the reconstructed time series.

Following identification of the significant components, median vectors for each were computed to capture the dominant periodic behavior identified through VBPBB. These median vectors were then integrated into the temperature dataset as auxiliary variables to support subsequent imputation using the Amelia II package. In this approach, the significant periodic components provided additional covariate information, allowing Amelia II to align imputations more closely with natural data fluctuations and preserve underlying temporal patterns. This integration was not a separate imputation method but was incorporated directly into the VBPBB-enhanced Amelia II framework. To evaluate the impact of incorporating periodic information, we compared two conditions: (1) Amelia II imputation using auxiliary periodic components derived from VBPBB, and (2) a baseline Amelia II imputation without periodic components. This comparative design allowed for a direct assessment of the added value of periodic information in improving data accuracy.

The effectiveness of each imputation approach was quantified by calculating Root Mean Square Error (RMSE) and Mean Absolute Error (MAE) against actual observed data. Our findings reveal that Amelia Imputation with VBPBB consistently resulted in lower values of Root Mean Square Error (RMSE) and Mean Absolute Error (MAE), with RMSE decreasing significantly from 10.353441to 4.629906, and MAE from 3.014825to 1.326239. This substantial reduction in error rates underscores the enhanced capability of VBPBB to handle the complexities associated with structured, non-random missing data scenarios. The integration of VBPBB appears to effectively capture and reconstruct the underlying periodic trends that are often distorted or obscured when traditional imputation methods are applied without such enhancements.

| Imputation Method | RMSE | MAE |
|---|---|---|
| **Amelia Imputation with VBPBB** | 4.629906 | 1.326239 |
| **Amelia Imputation without VBPBB** | 10.353441 | 3.014825 |

**Figure 5: Comparison of RMSE and MAE for Temperature Data Imputation Using Amelia II with and without VBPBB**

Additionally, visualization of results was conducted using ggplot2, we generated line graphs to compare the distributions of original and imputed temperature values. This visual assessment was crucial to determine the fidelity of the imputation process and to identify any potential biases introduced.

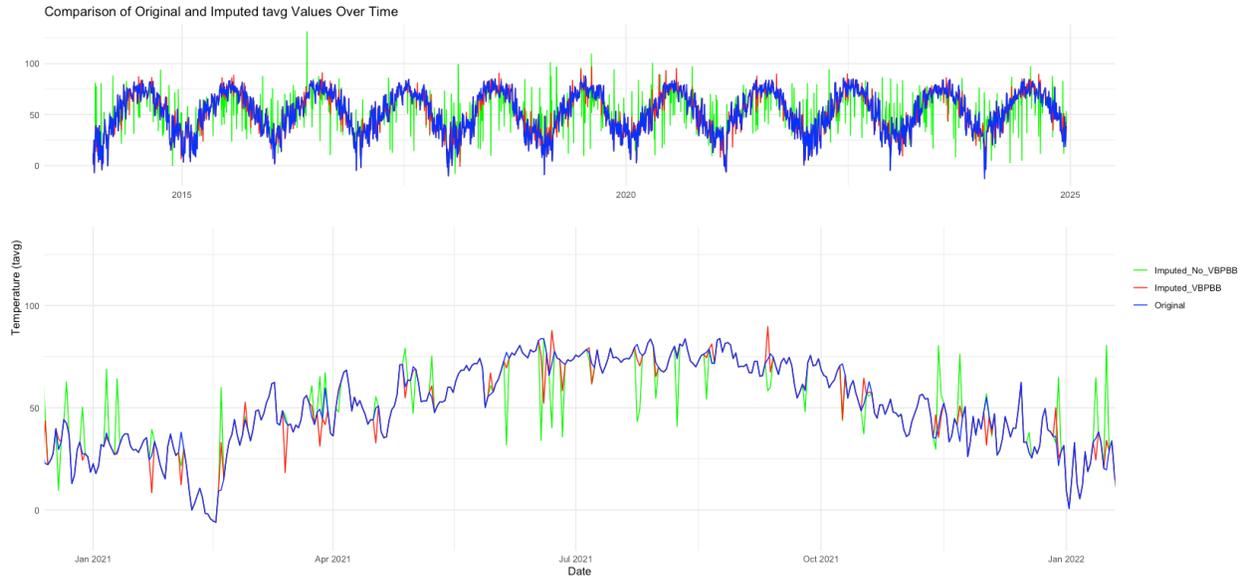

**Figure 6: Comparison of Original and Imputed Temperature Values Over Time**

This graph provides a comparative analysis of original and imputed daily average temperature (tavg) values over time, illustrating the effectiveness of different imputation methodologies. The original data (blue line) represents recorded temperature values, while the imputed values using the Variable Bandpass Periodic Block Bootstrap (VBPBB) method (red line) and those imputed without VBPBB (green line) highlight differences in how each approach preserves seasonal patterns and general trends. The top panel spans from 2014 to 2025, providing a long-term view of temperature fluctuations, while the bottom panel focuses on 2021, allowing for a closer inspection of short-term variations.

The VBPBB-enhanced imputation (red line) closely follows the original data (blue line), indicating that this method effectively maintains seasonal cycles and natural variability. In contrast, the imputed values without VBPBB (green line) exhibit greater deviations, particularly in capturing extreme values, suggesting distortions in temperature peaks and troughs. The zoomed-in section further highlights how VBPBB produces smoother and more reliable estimates, whereas the non-VBPBB imputation introduces larger fluctuations and inconsistencies.

The correlation analysis supports these observations, showing a stronger alignment between the VBPBB-imputed data and the original dataset (correlation = 0.9723945) compared to the non-VBPBB imputation (correlation = 0.8618343). This higher correlation confirms that VBPBB more accurately preserves periodic structures and seasonal trends, reducing estimation errors. In contrast, the lower correlation in the non-VBPBB imputation suggests a loss of critical periodic components, further reinforcing the superiority of VBPBB in time series data imputation. These findings demonstrate that VBPBB provides a more robust and reliable approach for handling missing values, ensuring greater accuracy in time-series analysis.

Also, we assessed the quality of the imputed data and the impact of subsequent smoothing techniques, we compare the original temperature data against the imputed and smoothed data using two different smoothing methods.

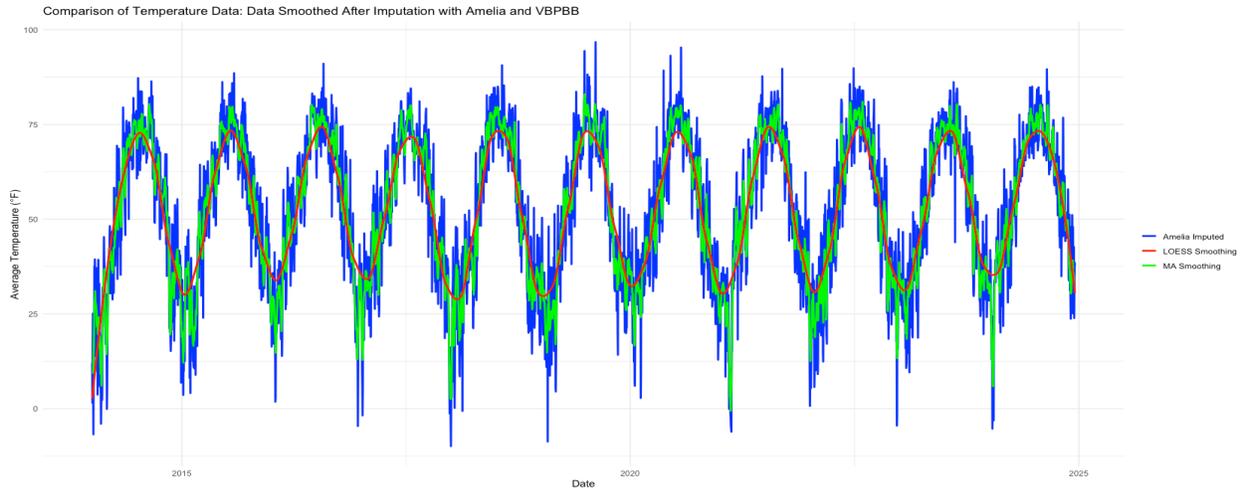

**Figure 7**: Temperature Data Smoothed After Imputation with Amelia and VBPBB (2014-2025)

This figure shows the original temperature data (blue line) alongside data smoothed using LOESS (red line) and MA (green line) techniques after imputation with VBPBB. The period from 2015 to 2025 highlights how VBPBB affects the imputation's fidelity to the original temperature trends.

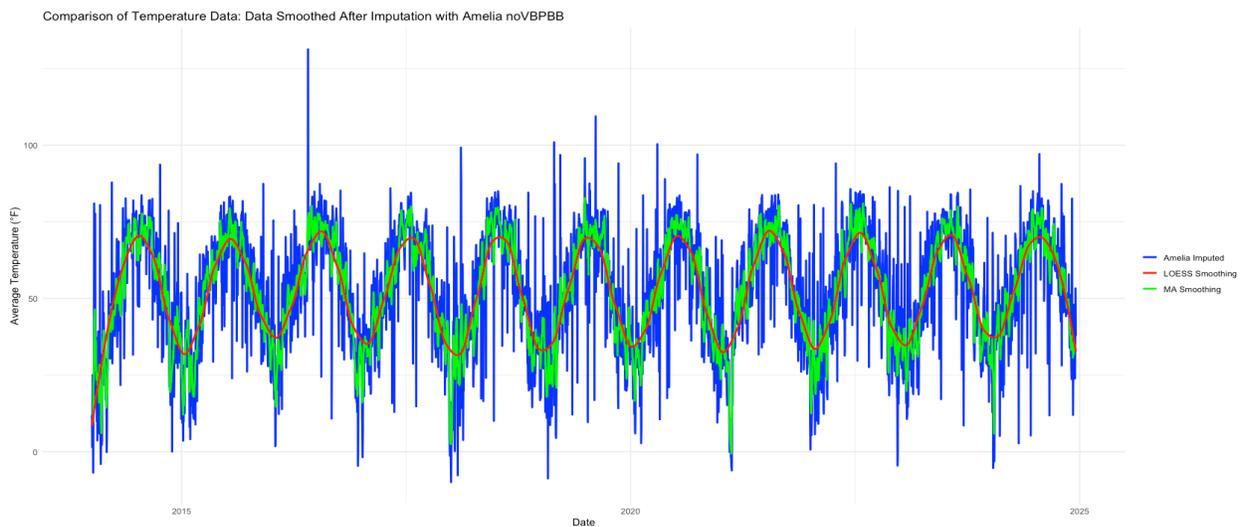

**Figure 8**: Temperature Data Smoothed After Imputation with Amelia without VBPBB (2014-2025)

This figure presents the original temperature data (blue line) against the LOESS smoothed (red line) and MA smoothed (green line) data after imputation without using VBPBB, illustrating the variance in data smoothing and trend following without the enhancement of VBPBB.

The application of smoothing techniques—specifically LOESS (Locally Estimated Scatterplot Smoothing) and Moving Average (MA)—serves primarily as a diagnostic and validation tool in this study. These methods are applied after the imputation process to evaluate the effectiveness of different imputation approaches in preserving meaningful temporal structures. Smoothing reduces short-term variability and highlights broader seasonal trends, enabling a clearer visual comparison between the imputed and original temperature series. By comparing smoothed curves from each imputation method—both with and without the Variable Bandpass Periodic Block Bootstrap (VBPBB)—to the known structure of the original data, we can assess how well the methods maintain long-term and periodic patterns. While smoothing does not improve the imputation itself, it plays a critical role in validating the integrity of the reconstructed time series by helping identify artifacts or distortions introduced during imputation. In this sense, smoothing supports a nuanced evaluation of imputation fidelity, particularly in capturing the expected climatic behaviors embedded in the data.

The results of this analysis reveal that smoothed outputs from VBPBB-imputed data exhibit more consistent alignment with seasonal trends compared to those imputed without VBPBB. Both LOESS and MA techniques help underscore these differences, affirming the utility of VBPBB in maintaining data integrity under conditions of structured missingness. These findings highlight the importance of selecting a robust imputation strategy, particularly for time series with known periodicity, and support the broader applicability of VBPBB in handling the complexities inherent in real-world datasets.

## 8. Discussion

In this study, we have rigorously evaluated the effectiveness of incorporating the Variable Bandpass Periodic Block Bootstrap (VBPBB) with Amelia II for imputing missing data, particularly in datasets characterized by significant seasonal and periodic variations. The use of VBPBB has demonstrated profound improvements in maintaining the integrity and statistical accuracy of the imputed data, which is essential for conducting reliable analyses in critical areas such as public health and epidemiology during ongoing global health crises like pandemics.

Integrating VBPBB with Amelia II not only refines the imputation process but also crucially preserves the statistical structure of the original dataset. This methodological enhancement enables more precise and dependable data analyses that are foundational for informed decision-making and effective policy formulation. Our empirical findings reveal a substantial enhancement in imputation accuracy, evidenced by a dramatic reduction in both Root Mean Square Error (RMSE) and Mean Absolute Error (MAE). Specifically, the application of VBPBB led to a 56% decrease in RMSE and a 57% decrease in MAE, underscoring significant improvements in the quality of imputed values and, consequently, the robustness of subsequent analyses in fields that rely heavily on accurate data representation.

Despite its advantages, the VBPBB method has several limitations that may impact its widespread implementation. Its application requires advanced statistical knowledge and significant computational resources, which could pose challenges for accessibility. The success of VBPBB largely depends on the presence of clear and distinguishable periodic patterns within the dataset; its effectiveness may diminish significantly in their absence, limiting its applicability in datasets lacking clear cyclic structures. Moreover, careful parameter selection for bandpass filters and bootstrap samples is critical, as any oversight in this process can compromise data accuracy. Another limitation is the requirement for extensive data to effectively identify and leverage periodic patterns, which may restrict its use in datasets with shorter time spans or smaller sample sizes. Addressing these limitations through methodological refinements and enhanced automation could improve the practicality and broader adoption of VBPBB in statistical analysis.

## 9. Conclusion

Given the complexity and resource demands of the VBPBB method, further investigation into its performance with higher proportions of missing data is essential. It is crucial to ascertain whether the observed benefits persist or even amplify as the volume of missing information increases. Future research should also extend the exploration of VBPBB's utility across various datasets and domains, such as environmental science and economics, where data accuracy is paramount. Additionally, a deeper examination of how VBPBB can be integrated with Multiple Imputation (MI) (VBPBB+MI) is necessary to evaluate its effectiveness in systematically improving imputation accuracy across different missing data mechanisms.

Investigating VBPBB+MI in scenarios with varying missing data proportions (MCAR, MAR, and MNAR) would provide insight into its robustness and potential limitations, ensuring its applicability in real-world settings. Extensive simulation studies should be conducted to assess its consistency in handling missing values under different periodic structures, levels of autocorrelation, and dataset sizes. Furthermore, integrating machine learning techniques with VBPBB+MI could democratize its application by automating elements of the imputation process, enhancing both accessibility and efficiency. To solidify its role as a valuable tool in statistical analysis, ongoing studies will need to validate and refine the application of VBPBB across diverse scenarios, ensuring that it consistently delivers high-quality imputations in a variety of contexts. This continued research will lay the groundwork for a future study focused on extensive simulations, further strengthening the statistical foundations of VBPBB+MI and its role in data-driven decision-making across critical fields.